\documentclass[prb,preprint]{revtex4}

\usepackage{amsmath}
\usepackage{graphicx}

\begin{document}

\title{Estimate of temperature and its uncertainty in small systems}
\author{M. Falcioni}
\affiliation{Dipartimento di Fisica, Universit\`a La Sapienza, p.\ le Aldo Moro 2, 00185 Roma, Italy}

\author{A. Puglisi}
\affiliation{ISC-CNR and Dipartimento di Fisica, Universit\`a La Sapienza, p.\ le Aldo Moro 2, 00185 Roma, Italy}

\author{A. Sarracino}
\affiliation{ISC-CNR and Dipartimento di Fisica, Universit\`a La Sapienza, p.\ le Aldo Moro 2, 00185 Roma, Italy}

\author{D. Villamaina}
\affiliation{Dipartimento di Fisica, Universit\`a La Sapienza, p.\ le Aldo Moro 2, 00185 Roma, Italy}

\author{A.Vulpiani}
\affiliation{Dipartimento di Fisica, Universit\`a La Sapienza, p.\ le Aldo Moro 2, 00185 Roma, Italy}


\begin{abstract}
The energy of a finite system thermally connected to a thermal reservoir may fluctuate, while the temperature is a constant representing a thermodynamic property of the reservoir. The finite system can also be used as a thermometer for the reservoir. From such a perspective the temperature has an uncertainty, which can be treated within the framework of estimation theory. We review the main results of this theory, and clarify some controversial issues regarding temperature fluctuations. We also offer a simple example of a thermometer with a small number of particles. We discuss the relevance of the total observation time, which must be much longer than the decorrelation time.
\end{abstract}

\maketitle

\section{Introduction}

In equilibrium thermodynamics there is a one-to-one relation between the energy of a macroscopic system, which is not at a phase transition, and its temperature. The role of the temperature is to control the transfer of energy between the system and other systems thermally coupled to it. Thermal (heat) reservoirs are assumed to have infinite energy and are characterized only by their temperature. A finite system in thermal contact with a thermal reservoir will attain the temperature of the reservoir, which is a device designed to bring a body to a well defined temperature.

If we know the energy $U$ of a thermodynamic system ${\cal A}$ in equilibrium, we can adopt two perspectives:

\begin{enumerate}

\item ${\cal A}$ plays the role of a thermometer and can be used to determine the temperature, $T(U)$, of a thermal reservoir ${\cal R}_T$ with which the system is, or had been, in contact.

\item ${\cal A}$ performs the role of a thermal reservoir and can be used to assign the same temperature $T(U)$ to all its subsystems. We may assume that ${\cal A}$ has been brought into contact with an appropriate reservoir to acquire the given $U$ and $T$ and then isolated, thus keeping its subsystems in equilibrium at the temperature $T$.
\end{enumerate}

The microscopic aspects of statistical mechanics alters these perspectives by introducing fluctuations of physical quantities in equilibrium so that a finite system in equilibrium with a thermal reservoir at temperature $T$ does not have a well defined energy, but a well definite distribution of energy, $P(E,T)$, and a well defined average energy, $\langle E \rangle = U(T)$. The temperature of the reservoir becomes the parameter that controls the distribution of the energy of the finite system.

Energy fluctuations are practically unobservable for macroscopic bodies. However, if systems of all sizes are considered, there is a conceptual problem with respect to these perspectives. If contact with the reservoir at temperature $T$ does not guarantee a unique energy of a system, but can determine only a distribution of energies for system, then if the isolated system ${\cal A}$ has a given energy, how can we be sure that ${\cal A}$ was in contact with reservoir ${\cal R}_T$ and not with reservoir ${\cal R}_{T'}$, with $T' \neq T$? To what extent can we assume that ${\cal A}$ and its subsystems are in equilibrium at temperature $T$ and not at temperature $T'$?

In statistical mechanics the (almost) one-to-one relation between $U$ and $T$ is recovered for macroscopic bodies, because of the relative smallness of the energy fluctuations. However, the problem of assigning a temperature to a given energy is relevant for non-macroscopic bodies.

Our understanding of temperature fluctuations has a long history. Einstein showed that the statistical properties of macroscopic variables can be determined in terms of quantities determined in thermodynamic equilibrium. In Sec.~\ref{review} we review the basic concepts of the Einstein approach. Temperature fluctuations have a special status in fluctuation theory. The Einstein theory yields formal expressions for $\langle (\delta T)^2\rangle$. Some authors suggest that temperature and energy are complementary, similar to position and momentum in quantum mechanics.\cite{UL99} In contrast, others have stressed the contradictory nature of the concept of temperature fluctuations\cite{K88}: in the canonical ensemble, which describes contact with a thermal reservoir, the temperature is a parameter, so it cannot fluctuate. For a discussion of temperature fluctuations see Refs.~\onlinecite{CSA92,K85}.

Mandelbrot has shown that the problem of assigning the temperature of the thermal reservoir, to which a system had been in thermal contact, can receive a satisfactory answer within the framework of estimation theory.\cite{M89} This analysis shows that as a system becomes smaller, the second perspective gradually loses its meaning (a small system cannot be considered as a thermal reservoir), and the first perspective maintains its validity, because a small system can be used as a thermometer by repeating the measurement of its energy a suitable number of times.

In the usual course on statistical physics the theory of fluctuations is explained using the energy or number of particles. Students might gain the impression that the same approach can be applied to other quantities such as the temperature. This issue is controversial, because temperature is usually a parameter and not a fluctuating quantity. With the advent of small systems such as nanosystems and biomolecules, a fluctuating temperature is often discussed in research.

Our paper is an effort to explain the possible pitfalls of generalizing fluctuation theory to the wrong quantities, as well as to illustrate a meaningful way of introducing temperature fluctuations. We first briefly review the contribution by Mandelbrot to understanding temperature fluctuations.\cite{M89} We will also discuss a model thermometer, which allows for a detailed understanding of the problem. We will see that, even in a system with few degrees of freedom, the temperature due to contact with a thermal reservoir is a well defined quantity that can be determined to arbitrary accuracy if a enough measurements are made. However, the observation time must be much longer than the decorrelation time of the underlying dynamics so that the number of independent measurements is sufficient.

The paper is organized as follows. In Sec.~\ref{review} we review the Einstein theory of fluctuations and discuss the origin of the problem. Section~\ref{mathstat} is devoted to a discussion on the relation between statistics and fluctuations (uncertainty) of the temperature. In Sec.~\ref{model} we present a model for a thermometer and illustrate a practical way to determine the temperature. In such a model, as well as for any thermometer, we have an indirect measurement of $T$, which is a statistical estimator, obtained by successive measurements of an observable.

\section{Review of the Einstein theory of fluctuations} \label{review}

For consistency we briefly recall the Einstein theory of fluctuations,\cite{LandauFisStat} focusing on the issue of temperature fluctuations.

Assume that the macroscopic state of a system is described by $n$ variables, $\alpha_1, \ldots,\alpha_n$, which depend on the microscopic state ${\bf X}$: $\alpha_j=g_j({\bf X})$, $j=1, \ldots, n$. Denote by ${\cal P}$ the parameters that determine the probability distribution function of the microscopic state ${\bf X}$. For example, in the canonical ensemble ${\cal P}_c = (T, V, N)$ and in the microcanonical ensemble ${\cal P}_m = (E, V, N)$. The probability distribution function of $\{\alpha_j \}$ is given by
\begin{equation}
P(\alpha_1, \ldots,\alpha_n)=\!\int\! \rho({\bf X},{\cal P}) \prod_{j=1}^n\delta(\alpha_j- g_j({\bf X}))\,d{\bf X},
\end{equation}
where $\rho({\bf X},{\cal P})$ is the probability distribution function of ${\bf X}$ in the ensemble with parameters ${\cal P}$. In the canonical ensemble we have
\begin{equation}
\label{eq-1} P(\alpha_1, \ldots,\alpha_n) = e^{- \beta\big[F(\alpha_1, \ldots,\alpha_n|{\cal P}_c) - F({\cal P}_c) \big]},
\end{equation}
where $\beta=1/k_BT$, $k_B$ is Boltzmann's constant, $F({\cal P}_c)$ is the free energy of the system with parameters ${\cal P}_c$, and $F(\alpha_1, \ldots,\alpha_n|{\cal P}_c)$ is the free energy of the system with parameters ${\cal P}_c$ and macroscopic variables $\alpha_1, \ldots,\alpha_n$:
\begin{equation}
F(\alpha_1, \alpha_2, \ldots,\alpha_n|{\cal P}_c)= -k_BT \ln \int \prod_{j=1}^n\delta(\alpha_j- g_j({\bf X})) e^{-\beta H({\bf X})} d{\bf X}.
\end{equation}
In the microcanonical ensemble we have
\begin{equation}
\label{prob-2} P(\alpha_1, \ldots,\alpha_n) = e^{ [S(\alpha_1, \ldots,\alpha_n|{\cal P}_m) - S({\cal P}_m)]/k_B} \equiv e^{\delta S(\alpha_1, \ldots,\alpha_n)/k_B},
\end{equation}
which is the Boltzmann-Einstein principle, where $S$ is the entropy.\cite{LandauFisStat}

For macroscopic systems it is natural to assume that the fluctuations with respect to thermodynamic equilibrium are small. Therefore we can expand $\delta S (\alpha_1, \alpha_2, \ldots,\alpha_n)$ in a Taylor series about the mean values $\{ \alpha_j \}$, which coincide with their values in thermodynamic equilibrium $\{ \alpha_j^{*} \}$:
\begin{equation}
\label{eq-3} \delta S (\alpha_1, \ldots,\alpha_n) \simeq -{1 \over 2} \sum_{i,j} \delta \alpha_i A_{ij} \delta \alpha_j
\end{equation}
where $\delta \alpha_j= \alpha_j - \alpha_j^{*}$, and
\begin{equation}
A_{ij}=-{\partial^2 S \over {\partial \alpha_j \partial \alpha_i}} \Big|_{{\mathbf \alpha}^{*}}.
\end{equation}
Therefore small fluctuations are described by a multivariate Gaussian probability distribution function:
\begin{equation}
P(\alpha_1, \ldots,\alpha_n) \simeq \sqrt{ {\mbox{det}\,{\bf A}} \over (2\pi k_B)^n} \, \exp \left\{-{1 \over 2k_B} \sum_{i,j} \delta \alpha_i A_{ij} \delta \alpha_j\right\},
\end{equation}
and
\begin{equation}
\label{eq-4} \langle\delta \alpha_i\delta \alpha_j\rangle=k_B\Bigl[{\bf A}^{-1} \Bigl]_{ij}.
\end{equation}
The entries of the matrix $A_{ij}$ are calculated at equilibrium. The matrix ${\bf A}$ must be positive (that is, all its eigenvalues must be strictly positive), which means that the difference of the entropy with respect to equilibrium must be negative. The well known expression for the energy fluctuations
\begin{equation}
\label{eq-5} \langle(E-\langle E\rangle)^2\rangle=k_BT^2C_V,
\end{equation}
where $C_V= \partial \langle E\rangle/\partial T$ is the heat capacity at constant volume, is a special case of Eq.~(\ref{eq-4}).

The $A_{ij}$ are functions of quantities evaluated at thermodynamic equilibrium, so that we can write $\delta S$ as a function of different variables. For instance, we can express $S$ as function of $T$ and $V$:\cite{LandauFisStat}
\begin{equation}
\label{eq-6} \delta S= - {C_V \over 2 T^2} (\delta T)^2 + {1 \over 2T} { \partial P \over \partial V}\Big|_T (\delta V)^2.
\end{equation}
By using Eqs.~(\ref{prob-2}) or (\ref{eq-4}), we obtain
\begin{equation}
\label{eq-7} \langle(\delta T)^2\rangle={k_B T^2 \over C_V}.
\end{equation}
Equation~(\ref{eq-6}) is correct if we consider $S$ as a state function. In contrast, Eq.~(\ref{eq-7}) follows from Eq.~(\ref{prob-2}) with $\delta S$ related to the probability distribution function of fluctuating quantities, and hence the derivation of Eq.~(\ref{eq-7}) is formal (in the sense of the mere manipulation of symbols) and its meaning is not clear. Note that in Eq.~(\ref{eq-6}) $\delta T=\delta (\partial E/\partial S)$.

For a system whose energy fluctuates about the value $\langle E \rangle$ (such that $\partial E/\partial S|_{E=\langle E \rangle} = T$, where $T$ is the temperature of the thermal reservoir) we can think of $\hat{T} \equiv \partial E/\partial S|_E $ as {the} temperature $ \hat{T} \neq T$ of this system, if it has been found with energy $E \neq \langle E \rangle$. However, we can also think of $\hat{T} $ as the best guess for $T$ if the energy $E$ has been measured. It is tempting to say that because temperature is proportional to the mean kinetic energy, its fluctuations are proportional to fluctuations of the kinetic energy. This pointis a delicate one, which will be considered in Sec.~\ref{conclude}.

If we assume Eq.~(\ref{eq-7}) and use Eq.~(\ref{eq-5}), we have
\begin{equation}
\label{eq-8} \langle(\delta T)^2\rangle\langle(\delta E)^2\rangle=k_B^2T^4 \quad \textrm{or} \quad \langle(\delta \beta)^2\rangle\langle(\delta E)^2\rangle=1.
\end{equation}
Equation~\eqref{eq-8} can be interpreted as a ``thermodynamic uncertainty relation'' formally similar to the Heisenberg principle. Some authors discuss a ``thermodynamic complementarity'' where energy and $\beta$ play the role of conjugate variables\cite{UL99} (see Sec.~\ref{uncert}).

Other authors, such as Kittel,\cite{K73,K88} claim that the concept of temperature fluctuations is misleading. The argument is simple: temperature is just a parameter of the canonical ensemble, which describes the statistics of the system, and therefore it is fixed by definition.

Some authors wonder about the meaning of the concept of temperature in small systems.\cite{S95} For instance, Feshbach\cite{F88} considered that for an isolated nucleus consisting of $N=O(10^2)$ nucleons (neutrons and protons), we expect from Eq.~(\ref{eq-7}) a non-negligible value of $\delta T/T$. However, from experimental data we observe (in Feshbach's words) that the empirical parameter to be identified with $\beta$ ``does not have such a large uncertainty.'' McFee\cite{MF73} wrote that ``The average temperature of a small system of constant specific heat connected to a thermal reservoir turns out to be different from that of the reservoir,'' and considered the fluctuations of $\beta(E)=\partial S(E)/\partial E$. Because such a quantity is a function of energy, its fluctuations are well defined and can be studied. He found
\begin{equation}
\langle(\delta \beta)^2\rangle= {\langle(\delta E)^2\rangle \over C_V^2 k_B^2 T^4},
\end{equation}
which is equivalent to Eq.~(\ref{eq-8}).

\section{Statistics and statistical mechanics} \label{mathstat}

In this section we illustrate the approach of Mandelbrot~\cite{M89} to statistical mechanics and review some basic concepts of statistics.

\subsection{Thermal Reservoirs}

A thermal reservoir is a system with very large (practically infinite) energy, such that a system with finite energy which is put in thermal contact with the reservoir comes to equilibrium at the temperature $T$ of the reservoir. In thermodynamics we consider only macroscopic bodies, which are those that have a well defined macroscopic energy by being in thermal equilibrium with a reservoir. Even in a purely phenomenological context we can discuss the fluctuations of the energy of a generic system in thermal equilibrium.\cite{S25,M89} However in statistical mechanics we can also consider systems with a few degrees of freedom, and therefore the fluctuations of the energy can be significant. The distribution of the energy $E$ of a system that is in equilibrium with a thermal reservoir of temperature $T$ is given by the Boltzmann-Gibbs density function
\begin{equation}
\label{gibbs} P (E, T) = \frac{G(E) \exp({-E/k_B T)}} {Z (T)},
\end{equation}
where $G(E)$ is the density of states and $Z(T)$ is the partition function.

When a system is in equilibrium with an thermal reservoir, we have two mutually exclusive situations: either we know the temperature of the reservoir and can describe the energy distribution of the system; or we do not know the temperature of the reservoir, and can determine it from the energy distribution of the system. The latter situation is called the inverse problem. For the inverse problem we can use the tools of estimation theory, which makes it possible to use the available data (in this case a series of energy values) to evaluate an unknown parameter (in this case $T$).

If we assume that the equilibrium properties of an isolated system, whether it has been isolated from a thermal reservoir or not, are described by the microcanonical probability density, an answer to the inverse problem is also an answer to the question: is it possible to assign a temperature to an isolated system with a given energy? The origin and the importance of the question resides in the following considerations. For an isolated system composed of $N$ non-interacting subsystems, we can calculate average values of observables by means of the probability density
\begin{equation}
\label{micro} \frac {1}{G_0(E)} g(u_1) \cdots g(u_{N-1}) g\!\left(u_N = E - \sum_1^{N-1} u_i \right),
\end{equation}
where $u_i$ is the energy of subsystem $i$, $E = \sum_1^N u_i$ is the energy of the system, and $g(u)$ is the energy density of a single subsystem, so that
\begin{equation}
\label{GdU} G_0(E) = \!\int\! g(u_1) \cdots g(u_N) \delta\!\left(\sum_1^N u_i - E \right) du_1 \cdots du_N.
\end{equation}
Such calculations are usually very difficult. However, for systems with a large number of subsystems\cite{Khinchin} we can approximate the probability density by a product of factors from the canonical ensemble:
\begin{equation}
\prod_i \frac {g(u_i) \exp({- u_i/k \widetilde{T})} } {Z_i (\widetilde{T})},
\end{equation}
where $\widetilde{T}$ is the temperature associated with the variable $E$. We have replaced non-independent variables by independent ones and have replaced $E$, the energy of the isolated system (and the parameter of the original distribution), by the common temperature (and the parameter of the approximate distributions) of its subsystems in thermal equilibrium. Note that this question must be posed in statistical mechanics, while in thermodynamics the functional relation between energy and temperature is an equation of state and does not call for a microscopic explanation.

\subsection{\label{estimation}Estimation Theory}

We recall here a few basic concepts from estimation theory.\cite{C46,K93} Consider a probability density function $f (x, \beta)$ of the variable $x$, which depends on the parameter $\beta$, together with a sample of $n$ independent events $(x_1, \dots, x_n)$, governed by the probability density $f$, so that the probability density of the sample is
\begin{equation}
\label{like} L (x_1, \dots, x_n, \beta) = f (x_1, \beta) \cdots f (x_n, \beta).
\end{equation}
We would like to estimate the unknown parameter $\beta$ from the values $\{x_i\}$. For this purpose we have to define a suitable function of $n$ variables, $\widehat{\beta}(x_1, \dots, x_n)$, to obtain the estimate of $\beta$ from the available information. The quantity $\widehat{\beta}$ is, by construction, a random variable. We can calculate, for instance, its expected value and its variance. We assume that $\widehat{\beta}$ is an unbiased estimate of $\beta$, that is, $\langle \widehat{\beta} \rangle = \beta$. It is clear that the usefulness of an estimating function is tightly linked to its variance.

Once the function $\widehat{\beta}(x_1,\dots,x_n)$ has been introduced, each sample $(x_1,\dots,x_n)$ can also be specified by giving the value of $\widehat{\beta}$ for the particular sample and the values of $n-1$ other variables $\lbrace \xi \rbrace$ that are necessary to specify the point on a surface of constant $\widehat{\beta}$. In other words, a change of variables $(x_1,\dots,x_n)\to(\widehat{\beta},\xi_1,\dots,\xi_{n-1})$ can be made, so that the probability of a sample may be written as
\begin{equation}
\label{prob} L (x_1, \dots, x_n)\,dx_1 \cdots dx_n = F (\widehat{\beta}, \beta) h(\xi_1, \dots, \xi_{n-1} \vert \widehat{\beta}, \beta)\,d\widehat{\beta}\,d\xi_1 \cdots d\xi_{n-1},
\end{equation}
where $F (\widehat{\beta}, \beta)$ is the density of the variable $\widehat{\beta}$ (depending on $\beta$) and $h(\lbrace \xi_i \rbrace \vert \widehat{\beta}, \beta)$ is the density of the variables $\lbrace \xi_i \rbrace$, conditioned by the value of $\widehat{\beta}$ and, in general, depending on the parameter $\beta$.

Given certain general conditions of regularity, we can obtain the Cram\'er-Rao inequality\cite{C46} for unbiased estimators:
\begin{equation}
\label{cramer-rao} \int\!\Big( \widehat{\beta} - \beta \Big)^2 F(\widehat{\beta}) d \widehat{\beta} \geq \left \{ n\!\int\! \Big(\frac{\partial}{\partial \beta} \ln f(x,\beta)\Big)^2 f(x,\beta) dx \right \}^{-1},
\end{equation}
where the denominator on the-right hand side of Eq.~\eqref{cramer-rao} is known as the Fisher information,\cite{C46} which gives a measure of the maximum amount of information we can extract from the data about the parameter to be estimated. This inequality puts a limit on the ability of making estimates, and also suggests that the estimator should be chosen by minimizing the inequality. When the variance of $\widehat{\beta}$ is the theoretical minimum, the result $\widehat{\beta}$ is an ``efficient estimate.''\cite{C46} We have followed the convention of distinguishing between an efficient estimate, which has minimum variance for finite $n$, and an asymptotically efficient estimate, which has minimum variance in the limit $n \to \infty$.

Starting from the probability of a given sample, Eq.~(\ref{like}), the method of maximum likelihood estimates the parameter $\beta$ as one that maximizes the probability, or is a solution of the equation
\begin{equation}
\label{max} \frac{\partial}{\partial \beta} \ln L (x_1, \dots, x_n, \beta) = 0.
\end{equation}
Under certain general conditions on some derivatives of $f(x,\beta)$ with respect to $\beta$,\cite{C46} Eq.~(\ref{max}) has a solution that converges to $\beta$ as $n \to \infty$. The solution is asymptotically Gaussian and is an asymptotically efficient estimate of $\beta$. In other words, there exists a random variable $\widehat{\beta} (x_1, \dots, x_n)$ which is a solution of Eq.~(\ref{max}) such that a maximum likelihood estimator of $\beta$ is obtained, whose probability density in the limit $n \to \infty$ approaches a normal probability density centered about $\beta$ with variance
\begin{equation}
\left \{ n \!\int \Big(\frac{\partial}{\partial \beta} \ln f(x,\beta)\Big)^2 f(x,\beta) dx \right \}^{-1}.
\end{equation}

\subsection{Thermal Reservoirs Again}

We now return to a system in equilibrium with a reservoir of unknown temperature on which we have performed a measurement of energy: the system considered in this section is a gas of $N$ classical particles. For simplicity, we begin by considering measurements of the energy $u$ of a single particle, whose probability distribution we write as
\begin{equation}
\label{gibbs-2} P (u, \beta) = \frac {g(u) \exp(-\beta u)} {Z (\beta)},
\end{equation}
where the parameter $\beta$ is $1/k_B T$ and the density of single particle states $g(u)$ is assumed to be known. Suppose that we have measured $n$ independent values of particle energy $(u_1, \dots, u_n)$. We can write
\begin{equation}
\label{gibbs-n} P (u_1, \dots, u_n, \beta) = \frac {g(u_1) \exp(-\beta u_1)} {Z (\beta)} \cdots \frac {g(u_n) \exp(-\beta u_n)} {Z (\beta)},
\end{equation}
or \begin{align} P (u_1, \dots, u_n, \beta) & = \frac {g(u_1) \cdots g(u_n) } {G_0(U)} \frac {G_0(U) \exp(-\beta U)} {Z^n (\beta)} \\ & \equiv h (u_1, \dots, u_{n-1} \vert U) P(U, \beta),
\end{align} where $U = \sum_1^n u_i$ and $G_0(U)$ is defined as in Eq.~(\ref{GdU}) with $N$ replaced by $n$; note that in this section $U$ has a different meaning with respect to the introduction. Because $P(U, \beta)$ is the probability density of measuring a total energy $U$ in $n$ independent single particle energy measurements, we see that $h (u_1, \dots, u_{n-1} \vert U)$, which is the conditional distribution of the energy in the sample given the total measured energy, does not depend on $\beta$. We conclude that good estimators of $\beta$ can be constructed as a function of the sum of the measured energies.

One possible choice of an estimator is the maximum likelihood estimator for which the value of $\widehat{\beta}$ is determined by
\begin{equation}
\label{mle-micro} - \frac {\partial }{\partial \beta} \ln Z^n (\beta) \Big \vert_{\widehat{\beta}} = \sum_1^{n} u_i.
\end{equation}
Equation~(\ref{mle-micro}) establishes a one-to-one relation between $\widehat{\beta}$ and $ \sum_i u_i $. For large $n$, the values of $\widehat{\beta}$ extracted from Eq.~(\ref{mle-micro}) are normally distributed around the true value $\beta$, with the variance
\begin{equation}
\label{eq26} \left \{ n \!\int\! \Big( u - \langle u \rangle \Big)^2 P (u, \beta) du \right \}^{-1} = \frac {1} {n \sigma^2_u},
\end{equation}
where
\begin{equation}
\langle u \rangle = \!\int\! u P (u, \beta)\,du,
\end{equation}
and $\sigma^2_u$ is the variance of the single-particle energy calculated with the true $\beta$.

For instance, if the density of states is $g(u) \propto u^{\eta}$, then from Eq.~(\ref{a1}) the maximum likelihood estimate is
\begin{equation}
\label{beta-mle} \widehat{\beta}_{\rm MLE} = \frac{n (\eta +1)}{U},
\end{equation}
which is not an unbiased estimate because as shown in the Appendix [see Eqs~(\ref{app-mle} and \ref{app-mle-1})]
\begin{equation}
\label{mean-mle} \big \langle \widehat{\beta}_{\rm MLE} \big \rangle = \beta \bigg ( 1 + \frac {1} {n (\eta +1) -1} \bigg) \neq \beta.
\end{equation}
As Eq.~(\ref{mean-mle}) shows, the maximum likelihood estimator is asymptotically unbiased and, from general theorems on maximum likelihood estimator [see the end of Sec.~\ref{estimation} and Eq.~(\ref{eq26})], we know that, because $\sigma^2_u = (\eta +1) \beta^{-2}$, we have for large $n$ \begin{align} \label{sigma-micro} \sigma^2_{\widehat{\beta}_{\rm MLE}} & \approx \frac {\beta^2} {n (\eta + 1)} \\ \noalign{\noindent or} \frac {\sigma_{\widehat{\beta}} }{\beta} & \approx \frac {1} {\sqrt {n (\eta + 1)}}.
\end{align} Therefore we can obtain an estimate of the parameter $\beta$ as accurate as we want by using a sufficiently large sample.

Another estimator for $\beta$ is given by the random variable
\begin{equation}
\label{beta-max} \widehat{\beta}_G = \frac {\partial }{\partial U} \ln G_0(U),
\end{equation}
where $G_0(U)$ is defined in Eq.~(\ref{GdU}) with $N$ replaced by $n$. Unlike the maximum likelihood estimator, the right-hand side of Eq.~\eqref{beta-max} is an unbiased estimator of $\beta$ for any $n$, but like the maximum likelihood estimator, it is not an efficient estimator for finite $n$.

For instance, with the density of states given as before, the estimate is (see Eq.~(\ref{GdU-app}) and the following discussion)
\begin{equation}
\label{beta-G} \widehat{\beta}_{G} = \frac{n (\eta +1) - 1}{U},
\end{equation}
with the variance
\begin{equation}
\label{sigma-G} \sigma^2_{\widehat{\beta}_{G}} = \beta^2 \bigg ( \frac {1} {n (\eta + 1) - 2} \bigg) > \frac {1} {n \sigma^2_u}.
\end{equation}
For this particular density of states Eq.~\eqref{sigma-G} also shows that $\widehat{\beta}_{G}$ becomes asymptotically efficient, because it attains the Cram\'er-Rao lower bound in the limit $n \to \infty$. This behavior is more general: we can demonstrate that for certain regularity conditions, these two estimators are asymptotically equivalent.\cite{S73,P77}

An important point of the preceding discussion is that, due to the exponential form of the canonical ensemble probability density, all of the information about $\beta$ is contained in the total energy of an isolated sample. We gain nothing by knowing the distribution of this energy among the $n$ elements of the sample. We say that $U= \sum_1^n u_i$ is sufficient for estimating $\beta$. Therefore we may also argue as follows.

Instead of $n$ measurements of the molecular energy, we make one measurement of the energy $E$ on the macroscopic system with density $P(E, \beta) = G(E) \exp(-\beta E)/Z_N (\beta)$. $G(E)$ is the density of states of the entire system, which reduces to $G_0(E)$ for systems made of non-interacting components. The Cram\'er-Rao inequality becomes
\begin{equation}
\label{CR-macro} \int\! \Big( \widehat{\beta} - \beta \Big)^2 F(\widehat{\beta}) d \widehat{\beta} \geq \frac {1} {\sigma^2_E},
\end{equation}
where $\sigma^2_E$ is the variance of the canonical energy of the macroscopic body.

For an ideal gas of $N$ identical particles, $\sigma^2_E = N \sigma ^2_u$, and Eq.~\eqref{CR-macro} becomes $\sigma^2_{\widehat{\beta}} \geq 1/N \sigma^2_u$. With regard to the determination of $\beta$, a single value of the macroscopic energy contains the same information as $N$ microscopic measurements.

We know that a non-ideal gas of $N$ identical particles with short-range interparticle interactions behaves (if not at a phase transition) as if it were composed of a large number, $N_{\rm eff} \propto N$, of (almost) independent components, and $\sigma^2_E \approx N_{\rm eff}\,\sigma^2_c$, where $\sigma ^2_c$ is the variance of one component. For instance, consider a system of $N$ particles in a volume $V$ with a correlation length $\ell = (c V/N)^{1/3}$, where $c\gg 1$ indicates strong correlations. We have $N_{\rm eff}\sim V/\ell^3= c^{-1} N$. Thus, even if $n=1$ in Eq.~(\ref{CR-macro}), that is, we perform a single measurement of energy, the variance of $E$, which is the energy of a macroscopic system, is extensive and the variance of $\widehat{\beta}$ may be small. We have $\sigma^2_{\widehat{\beta}} \geq 1/N_{\rm eff}\, \sigma^2_c$, with $N_{\rm eff} \propto N \gg 1$. By looking at $E$ as the result of $N_{\rm eff}$ elementary energy observations, our preceding considerations can be applied here with $N_{\rm eff}$ playing the role of $n$. In particular, the asymptotic properties for large $N_{\rm eff}$ of the two estimators are preserved, and the estimates of $\beta$ obtained by the two expressions \begin{subequations} \label{mle-E} \begin{align} & - \frac {\partial }{\partial \beta} \ln Z_N (\beta) \Big \vert_{\widehat{\beta}_{\rm MLE}} = E,\\ \noalign{\noindent and} & \widehat{\beta}_G = \frac {\partial }{\partial E} \ln G(E)
\end{align}
\end{subequations} approach the same value for $N_{\rm eff} \gg 1$, a condition that is verified for macroscopic bodies. Therefore, for a macroscopic system, we can obtain a good estimate of $\beta$ even with a single measurement of its energy, and we can assign a reliable value of $\beta$ to an isolated macroscopic system.

We have given an estimation theory justification of the standard definition of the temperature in statistical mechanics either in the canonical or microcanonical ensemble by means of Eq.~(\ref{mle-E}).

\subsection{\label{uncert}Uncertainty relations in statistical mechanics?}

From our discussion we see that the fluctuations of the random variables $\widehat{\beta}_{\rm MLE}$ and $\widehat{\beta}_{G}$ when $n \gg 1$ are approximately Gaussian with a variance $1/(n \sigma^2_u)$. The fluctuations of the total energy of the sample $U = \sum_1^n u_i$ also become Gaussian (by the central limit theorem) with variance $n \sigma^2_u$. Therefore, in this limit, we have $\sigma^2_{\widehat{\beta}}\,\sigma^2_U = 1$. Is there a deeper meaning? As can be seen, for instance, for $g(u) \propto u^{\eta}$, from Eqs.~(\ref{beta-mle}) and (\ref{beta-G}), $\widehat{\beta}_{\rm MLE}$ and $\widehat{\beta}_{G}$ are functions of $U/n$, and therefore, from $\sigma^2_U \propto n $ $\sigma^2_{\widehat{\beta}} \propto \sigma^2_U/n^2 \sim 1/n$.

Although the Rao-Cram\'er inequality, Eq.~(\ref{CR-macro}), is formally similar to Eq.~(\ref{eq-8}), which was obtained in the framework of Einstein's theory, the analogy is inexact and misleading. In mathematical statistics the quantity $\sigma^2_{\widehat{\beta}}$ measures the uncertainty in the determination of the value of $\beta$ and not the fluctuations of its values.

\section{\label{model}Model Thermometer}

To illustrate the ideas we have discussed in we discuss the following mechanical model for a thermometer. A box is filled with $N$ non-interacting particles of mass $m$. On the top of the box is a piston of mass $M$ which can move without friction in the $\hat{x}$ direction. Although the box is three-dimensional, only the motion in the $\hat{x}$ direction is relevant because we assume that the particles interact only with the piston. The other directions are decoupled from $\hat{x}$, independently of their boundary conditions. The one-dimensional Hamiltonian of the system is
\begin{equation}
{\cal H}=\sum_{i=1}^N\frac{p_i^2}{2m}+\frac{P_M^2}{2M}+FX,
\end{equation}
where $X$ is the position along the $\hat{x}$ axis of the piston, and the positions of the particles $x_i$ along the same axis are constrained to be between 0 and $X$. A force $F$ acts on the piston, and in addition there are elastic collisions of the gas particles with the piston. The particles exchange energy with a thermostat at temperature $T$ placed on the bottom of the box at $x=0$. When a particle collides with the ground, it acquires a speed $v$ with probability density\cite{Banavar}
\begin{equation}
P(v) = \ \frac{m}{k_B T} v\,e^{-mv^2/2 k_B T}.
\end{equation}
In the following we set $k_B=1$, which is equivalent to measuring the temperature in units of $1/k_B$.

\begin{figure}[h!]
\includegraphics[clip=true,width=8cm]{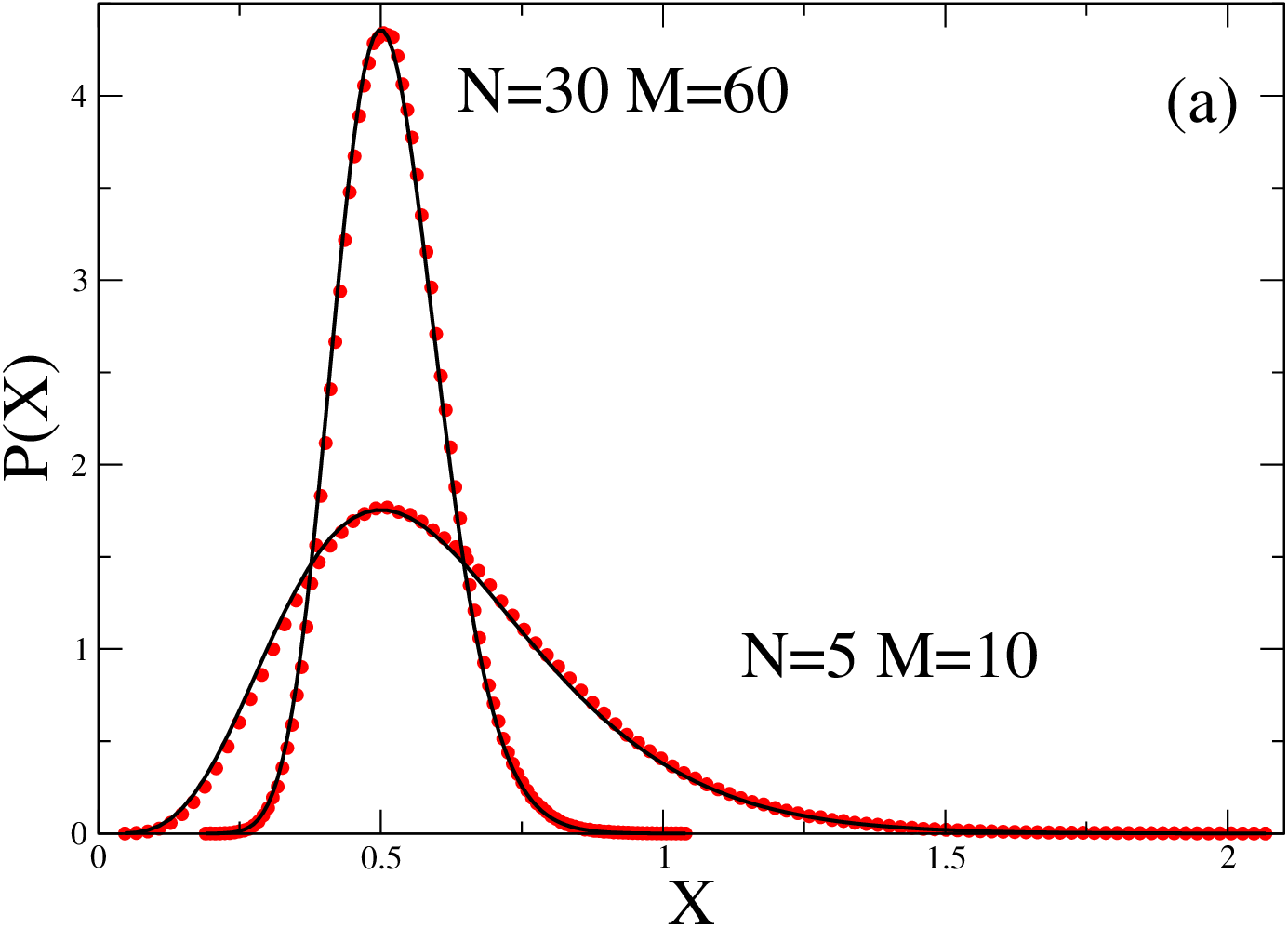}
\includegraphics[clip=true,width=8cm]{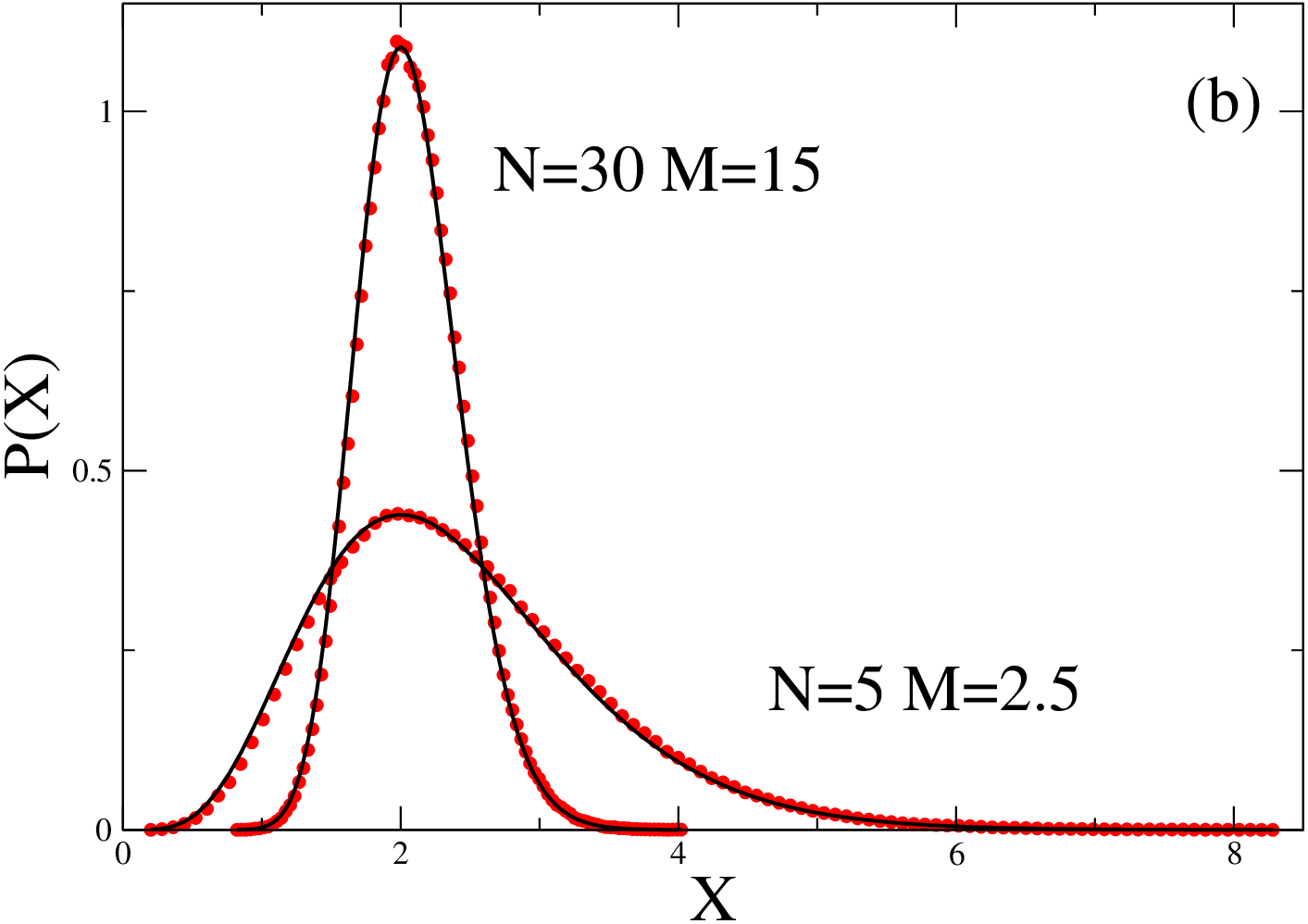}
\caption{(\label{fig_pdf}Color online). The probability distribution function of the position $X$ of the piston obtained by numerical simulations with $F=10$ and $T=1$ (dots) for different values of $N$ and $M$. (a) $N=30$ and $N=5$ with $M=2N$. (b) $N=30$ with $M=N/2$. The black lines show the analytical result from Eq.~(\ref{pdfX}). Each simulation has been performed up to a time such that each particle collided with the piston at least $10^5$ times.}
\end{figure}

The statistical mechanics of the system can be obtained in the canonical ensemble. The probability distribution function for the positions of the particles is
\begin{equation}
P(x_1,\ldots,x_N,X)=c_N\prod_{i=1}^N \theta(X-x_i)e^{-\beta FX},
\end{equation}
where $c_N=(\beta F)^{N+1}/\Gamma(N+1)=(\beta F)^{N+1}/N!$. We integrate over the positions of the particles and obtain
\begin{equation}
P(X)=\frac{1}{N!}(\beta F)^{N+1} X^N e^{-\beta FX}. \label{pdfX}
\end{equation}
The mean value $\langle X\rangle$ is
\begin{equation}
\langle X\rangle=\frac{(N+1)T}{F}. \label{aveposition}
\end{equation}
The probability distribution function for $X$ obtained by numerical simulations of the system is in agreement with Eq.~(\ref{pdfX}) even for small values of $N$, as can be seen in Fig.~\ref{fig_pdf}.

\begin{figure}[h!]
\includegraphics[clip=true,width=12cm]{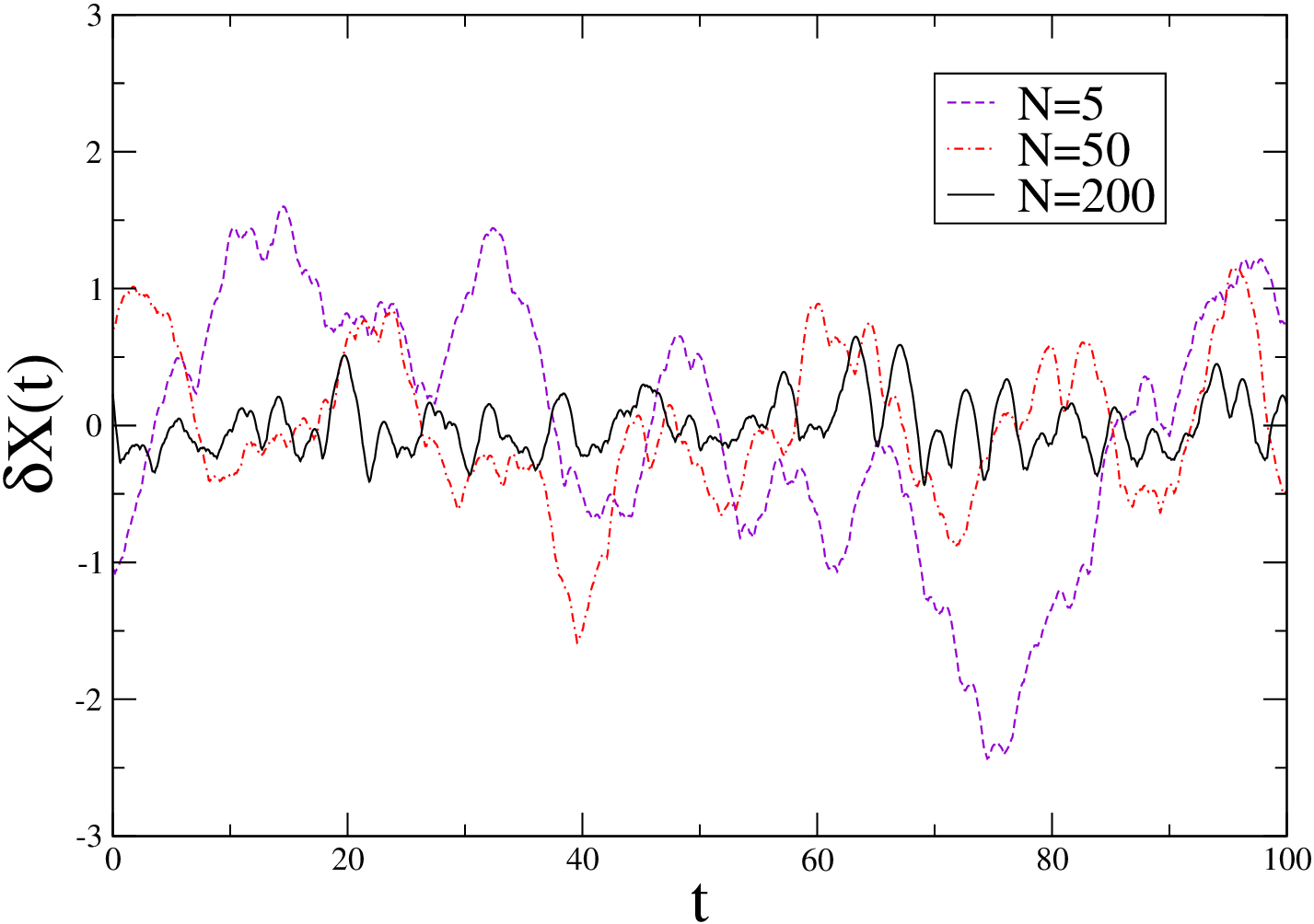}
\caption{\label{fig_X}(Color online). Time series of the displacement of the piston from its mean value for different values of $N$. The other parameters are $M=10$, $F=10$, $T=1$, and $m=1$. One can see that, although the qualitative behavior does not change with $N$, the amplitude of fluctuations decreases with $N$. }
\end{figure}

In the following we will estimate the temperature from a single (long) time series of the simulations of the piston position. This procedure is common both in numerical and in real experiments. Therefore it is necessary to consider the dynamical statistical properties of our system. For $N\gg 1$ and $M/m \gg 1$ we expect that the variable $\delta X\equiv X-\langle X\rangle$ is described by a stochastic process.\cite{H99} Figure~\ref{fig_X} shows the typical behavior of $\delta X(t)$ for different values of $N$. For our purposes (in particular, for $N$ not too large) it is not necessary to perform an accurate analysis.

We now discuss a measurement of the temperature with its uncertainty, regardless of the number of degrees of freedom. We assume that only the macroscopic degree of freedom, the position of the piston, is experimentally accessible. We want to determine the temperature and its uncertainty by a series of measurements. From Eq.~(\ref{aveposition}) the temperature can be estimated as
\begin{equation}
\hat{T}=\frac{F\hat{X}}{N+1},
\end{equation}
where $\hat{X}$ is an estimate of the average piston position. Assume that we have $\mathcal{N}$ independent measurements $X^{(1)},\ldots,X^{(\mathcal{N})}$. Because of the peculiar shape of the probability distribution function~(\ref{pdfX}) (it is an infinitely divisible distribution\cite{gned-kol}), we have that the variable $\hat{X}_{\mathcal{N}}= (X^{(1)}+\ldots +X^{(\mathcal{N})})/\mathcal{N}$ has a probability distribution function of the same shape, where $N$ is replaced by $N\mathcal{N}$ in Eq.~(\ref{pdfX}). The variance of $\hat{X}_{\mathcal{N}}$ is $\sigma^2_{\hat{X}}/\mathcal{N}$, because the values of $X$ are independent. From Eq.~\eqref{pdfX} we have $\sigma^2_{\hat{X}}=(N+1)/\beta^2 F^2$ [see the Appendix for the calculation of moments of the distribution~\eqref{pdfX}], and therefore
\begin{equation}
\sigma^2_{\hat{X}_{\mathcal{N}}}=\frac{1}{\mathcal{N}}\frac{N+1}{\beta^2 F^2}. \label{varposmean}
\end{equation}

An analysis of the distribution (\ref{pdfX}) shows that the Cram\'er-Rao lower bound for the estimators of $T$ is
\begin{equation}
\label{estimator-var} \frac{T^2}{\mathcal{N} (N+1)},
\end{equation}
so that we can verify that the random variable
\begin{equation}
\hat{T}=\frac{F}{N+1}\left(\frac{1}{\mathcal{N}}\sum_{i}X_{i}\right) \label{temp_estimator}
\end{equation}
is an unbiased and efficient estimator for every $\mathcal{N}$.

We now discuss how to determine the temperature and its uncertainty from a time series $\{X_{i}\}^{\mathcal{N}}_{i=1}$, where $X_i=X(i\delta t)$ and $\delta t$ is the sampling time interval; $\mathcal{N}\delta t$ is the total observation time. The procedure we will describe is also valid for non-independent data $\{X_{i}\}$, and depends only on the validity of Eq.~(\ref{aveposition}) and not on Eq.~(\ref{pdfX}).

\begin{figure}[h!]
\includegraphics[clip=true,width=12cm]{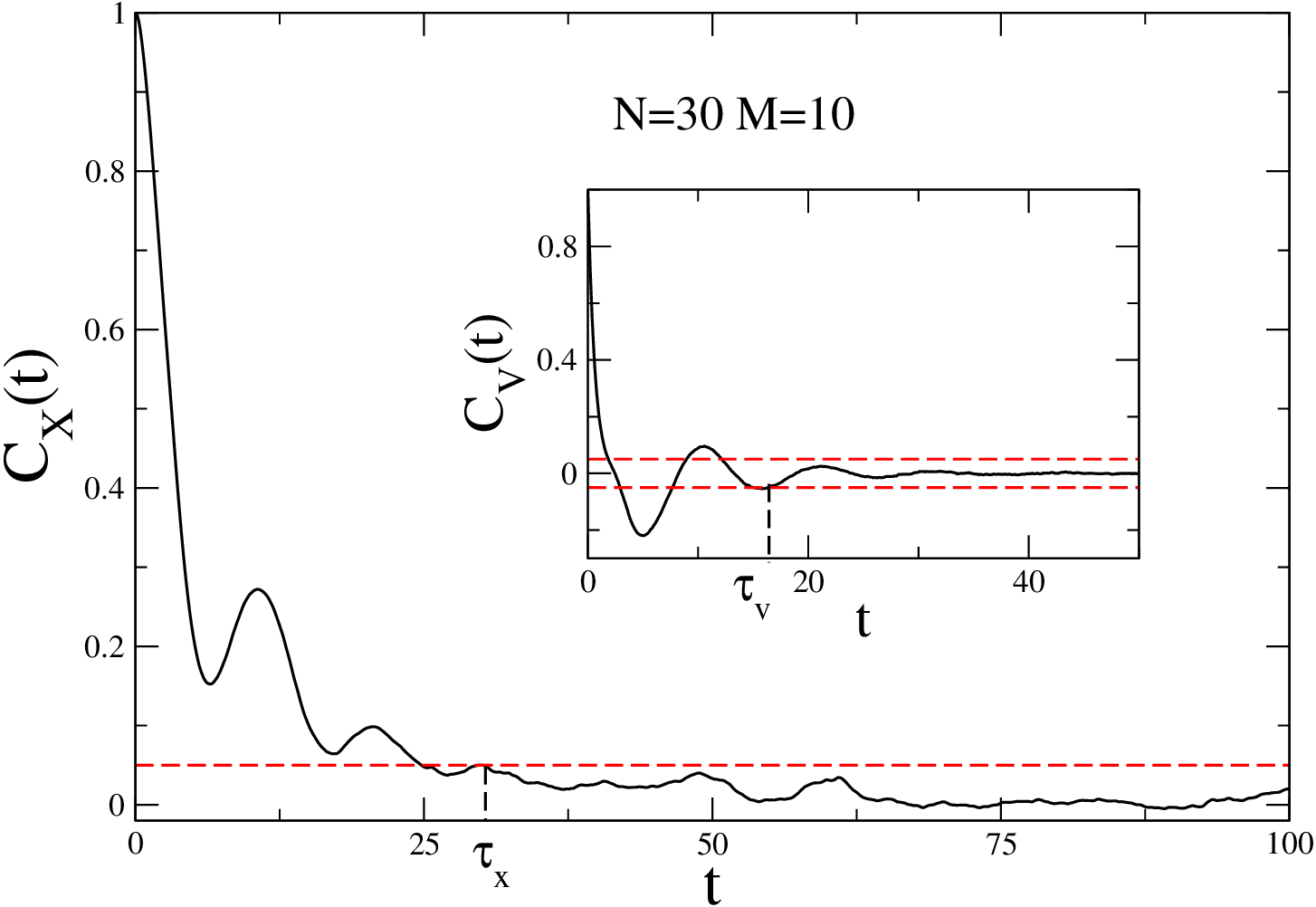}
\caption{\label{corrX}(Color online). The autocorrelation function of the position $X$ and velocity $V$ of the piston for $N=30$, $M=10$, $F=10$, $T=1$, and $m=1$. The estimates of the correlation times $\tau_X$ and $\tau_V$ are shown. Although the two correlation functions $C_X(t)$ and $C_V(t)$ are different, the corresponding characteristic times are of the same order.}
\end{figure}

\begin{figure}[h!]
\includegraphics[clip=true,width=12cm]{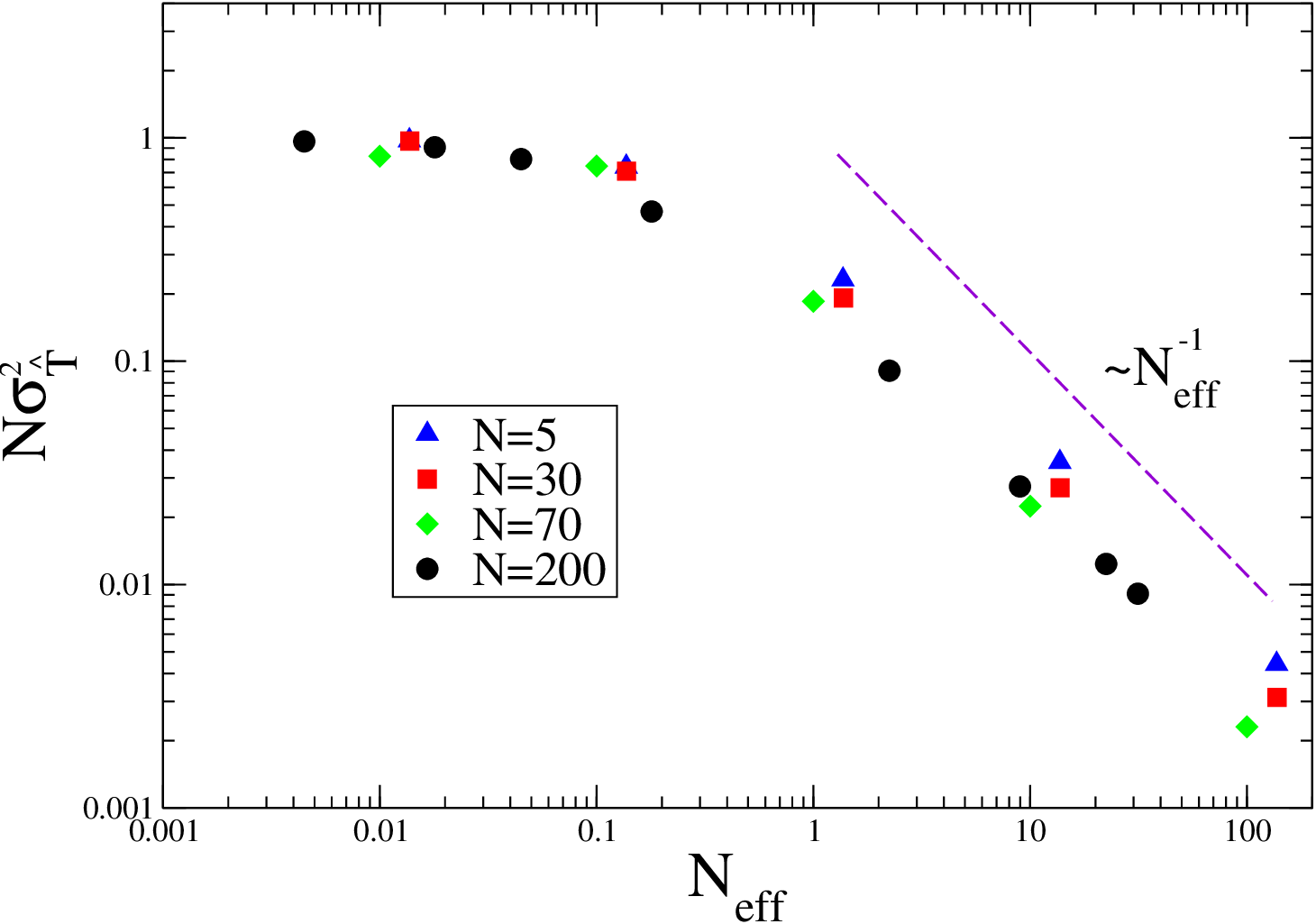}
\caption{\label{figT}The quantity $N \sigma^{2}_{\hat{T}}$ for different values of $N$ is numerically calculated and plotted as function of $\mathcal{N}_{\rm eff}=\mathcal{N}\delta t/\tau$. For large times, the uncertainty goes to zero as $1/\mathcal{N}_{\rm eff}$. The parameters are $\delta t=0.01$, $M=10$, $m=1$, $F=10$, $T=1$, $N=5$, 30, 70, and 200. It is clear that for the uncertainty of $T$, the relevant quantity is $N_{\rm eff}$ which depends both on $N$ and $\tau$.}
\end{figure}

The variance $\sigma^{2}_{\hat{T}}$ of the estimator $\hat{T}$, given by Eq.~(\ref{estimator-var}) is of order $\sim 1/N$, which can be non-negligible for single measurements on small systems. As described in Sec.~\ref{mathstat}, it can be arbitrarily reduced by increasing the number $\mathcal{N}$ of measurements. To clarify this point, we numerically computed the variance $\sigma^2_{\hat{T}}$ for several values of $N$ as a function of $\mathcal{N}$. In general, the data are correlated, and a correlation time $\tau$ must be estimated numerically. The simplest way is to look at the shape of the correlation functions of the observables of interest. If $\delta t < \tau$, the effective number of independent measurements is approximately $\mathcal{N}_{\rm eff}=\mathcal{N}\delta t/\tau$. By plotting $N\sigma^2_{\hat{T}}$ versus $\mathcal{N}_{\rm eff}$ we expect that the dependence on $N$ disappears, resulting in a collapse of the curves (see Fig.~\ref{figT}). Note that for large times, the uncertainty goes to zero as $1/\mathcal{N}_{\rm eff}$, in agreement with Eq.~(\ref{varposmean}). We will see that it is sufficient to know the typical time scales of the process. Namely, we need to determine the correlation functions $\langle\delta X(t)\delta X(0) \rangle$ or $\langle\delta V(t)\delta V(0)\rangle$, where $\delta V=\delta\dot{X}$. We can define a characteristic time for $X$, $\tau_{X}$, as the minimum time $t$ such that $\left\vert C_{X}(t) \right\vert< 0.05$, where
\begin{equation}
C_{X}(t)=\frac{\langle\delta X(t)\delta X(0) \rangle}{\langle\delta X^{2}(0)\rangle}.
\end{equation}
In the same way we can introduce the characteristic time for the velocity autocorrelation $\tau_{V}$ (see Fig.~\ref{corrX}).

The previous estimate of the temperature is well-posed once we know that Eq.~(\ref{aveposition}) holds. However, there are other possibilities. For instance we can choose to monitor the velocity of the piston instead of the position, and repeat the same analysis. It is straightforward to see that the velocity is Gaussian distributed, with variance $\left\langle V^{2}\right\rangle=T/M$, and only the mass of the piston is needed for the estimate. The choice of which estimator is more suitable is a matter of convenience.

For systems exhibiting aging and ergodicity breaking there are model thermometers that are very different from the one proposed here and are based on the linear response of the system and its comparison with unperturbed correlators.\cite{CKP97, NOI} In this case we may have access to ``effective'' temperatures related to slow, non-equilibrated, degrees of freedom, and the problem of uncertainty becomes more complicated, because the effective temperatures may be time-dependent.\cite{CKP97}

\section{\label{conclude}Conclusions}
We have considered the concept of temperature fluctuations and showed that it makes sense only when associated with uncertainties of measurement. In a molecular dynamics computation at fixed energy it is common practice to look at the fluctuations of the kinetic energy. Because the mean value of the kinetic energy per particle, $K$, is proportional to the temperature, it might be concluded that the fluctuations of $K$ are related to the fluctuations of the temperature, $\left\langle (\delta T)^2 \right\rangle \propto \left\langle (\delta K)^2 \right\rangle$. However, it is the random variable $K$ that fluctuates from sample to sample, not its mean value. Hence, in the preceding relation we should properly write $\left\langle (\delta \hat{T})^2 \right\rangle$ instead of $\left\langle (\delta T)^2 \right\rangle$, because we are using the random variable $K$ as an estimator of the temperature, which as stressed by Kittel\cite{K88,K73}, is an unknown but fixed parameter. If the conceptual difference between parameters and fluctuating variables is ignored, we may obtain suggestive but deceiving interpretations of results such as Eq.~(\ref{eq-8}). Estimation theory gives a precise role to uncertainties in temperature measurements, and establishes which properties make a temperature estimator better than others and leads to meaningful results such as Eq.~(\ref{cramer-rao}). In this framework we can appreciate the different use we can make of a system, either as a reservoir or as a thermometer. For the model system of Sec.~\ref{model} the relative uncertainty in the temperature estimation is
\begin{equation}
\frac{\Delta \hat{T}} {T} = \frac {\sqrt{\sigma^2_{\hat{T}}}}{T} = \frac{1}{\sqrt{\mathcal{N} (N+1)}}.
\end{equation}
If $N \gg 1$, even a single estimate, $\mathcal{N} = 1$, leads to a precise value for $T$. The system is not only an efficient thermometer, but also can be thought of as a reservoir with a definite temperature. In contrast, if $N$ is small, a single measurement can lead to a large uncertainty in the estimate of temperature, but the uncertainty can be arbitrarily reduced by increasing the number of measurements. The small system can be used as a thermometer, in this case the dynamics plays a non-negligible role, and the characteristic decorrelation time of the relevant variables dictates the effective number of independent measurements.

For systems with a few degrees of freedom the uncertainty in the temperature can be reduced by an adequate amount of data, which is an answer to the problem concerning $\beta$ pointed out by Feshbach in Ref.~\onlinecite{F88}.

\begin{acknowledgments}
The work of A.\ P.\ and A.\ S.\ is supported by the Granular-Chaos project, funded by the Italian MIUR under the FIRB-IDEAS grant number RBID08Z9JE.
\end{acknowledgments}

\appendix*

\section{\label{app}Some useful formulas}

To obtain Eq.~(\ref{beta-mle}) with $g(u) = \gamma u^{\eta}$ ($\gamma$ independent of $u$), we have
\begin{align}
Z (\beta) &= \!\int_0^{\infty} g(u) \exp (-\beta u)\, du = \int_0^{\infty} \gamma u^{\eta} \exp (-\beta u) \, du \\ &= \gamma \Big(\frac{1}{\beta}\Big)^{\eta + 1} \Gamma (\eta +1), \label{a1}
\end{align} where $\Gamma(z)$ (with $z > 0$) is the Gamma function. From Eq.~(\ref{mle-micro}) we immediately find Eq.~(\ref{beta-mle}).

We now derive Eqs.~(\ref{mean-mle}), (\ref{beta-G}), and (\ref{sigma-G}) starting from the power-law density of states $g(u)$. First, we show that for the density
\begin{equation}
P(U, \beta) = \frac {G_0(U) \exp(-\beta U)} {Z^n (\beta)} = \frac {G_0(U) \exp(-\beta U)} {\int_0^{\infty} G_0(U) \exp(-\beta U) d U},
\end{equation}
we have $G_0(U) \propto U^{n (\eta +1) -1}$. If we make the change of variables $u_i= x_i U$ in Eq.~(\ref{GdU}), we obtain
\begin{equation}
G_0(U) = \gamma^n U^{n(\eta +1)}\int_0^{\infty} \cdots \int_0^{\infty} x_1^{\eta} x_2^{\eta} \cdots x_n^{\eta} \frac {\delta (\sum_1^n x_i - 1)} {U} \, dx_1 dx_2 \cdots dx_n,
\end{equation}
where we have used the relation $\delta(aw) = \delta(w)/a$. We make the $U$ dependence explicit and write
\begin{equation}
\label{GdU-app} G_0(U) = \gamma^n U^{n(\eta +1) -1} I(n, \eta).
\end{equation}

We can calculate the averages
\begin{equation}
\label{ave} \Big \langle \frac{1} {U^k} \Big \rangle = \frac {\int\! U^{n(\eta +1) -1} (1/U^k) \exp (-\beta U) \, dU} {\int U^{n(\eta +1) -1} \exp (-\beta U) \, dU}
\end{equation}
by means of the integrals
\begin{equation}
\label{gamma} \int_0^{\infty}\! U^{\alpha} \exp (-\beta U) \, dU = \beta^{-(\alpha +1)} \Gamma (\alpha +1).
\end{equation}

From Eqs.~(\ref{beta-mle}), (\ref{ave}), and (\ref{gamma}) and the property $z \Gamma (z) = \Gamma (z+1)$, we obtain \begin{align} \label{app-mle} \big \langle \widehat{\beta}_{\rm MLE} \big \rangle &= n(\eta +1) \Big \langle \frac{1}{U} \ \Big \rangle = n(\eta +1) \beta \frac {\Gamma (n(\eta +1) - 1)}{\Gamma (n(\eta +1))} \\ \label{app-mle-1} &= n(\eta +1) \beta \frac {1}{n(\eta +1)-1},
\end{align} which is Eq.~(\ref{mean-mle}).

From Eqs.~(\ref{beta-max}) and (\ref{GdU-app}) we obtain Eq.~(\ref{beta-G}) and $\langle \widehat{\beta}_{G} \rangle = \beta$. From Eq.~(\ref{ave}) with $k=2$ we obtain
\begin{equation}
\label{A10} \Big \langle \frac{1}{U^2} \Big \rangle = \frac {\beta^2 } {(n (\eta +1) -1) (n (\eta +1) -2)}.
\end{equation}
By recalling the definition of $\widehat{\beta}_G$ in Eq.~(\ref{beta-G}), the fact that $\sigma^2_{\widehat{\beta}_{G}} = \langle \widehat{\beta}_{G}^2 \rangle - \langle \widehat{\beta}_{G}\rangle^2$, and $\langle \widehat{\beta}_{G} \rangle = \beta$, we arrive at Eq.~(\ref{sigma-G}).

With similar calculations involving the Gamma function, we arrive at Eq.~(\ref{varposmean}) for the variance of the variable $\hat{X}_{\mathcal{N}}= (X^{(1)}+\ldots +X^{(\mathcal{N})})/\mathcal{N}$, that is $\sigma^2_{X}/\mathcal{N}$. If Eq.~(\ref{pdfX}) gives the distribution of $X$, its moments are
\begin{equation}
\langle X^k\rangle=\frac{(\beta F)^{N+1}}{N!} \int_0^{\infty}X^{N+k}e^{-\beta FX}dX.
\end{equation}
In terms of the variable $z=\beta FX$, we obtain
\begin{equation}
\langle X^k\rangle=\frac{1}{N!(\beta F)^k}\int_0^{\infty}z^{N+k}e^{-z}dz =\frac{\Gamma(N+k+1)}{N!(\beta F)^k}=\frac{(N+k)\ldots(N+1)}{(\beta F)^k}.
\end{equation}
Therefore, $\sigma_X^2=\langle X^2\rangle-\langle X\rangle^2= (N+1)/(\beta F)^2$.

\section*{Figure Captions}

\end{document}